\documentstyle[12pt]{article}

\global\arraycolsep=1pt
   \def \dS{\delta \Sigma^k}
   \def \dO{\delta \omega^k}

\begin{document}

\date{TIT/HEP-224/COSMO-32 June, 1993}

\title{Moduli Space of Topological 2-form Gravity}

\author{Mitsuko Abe\thanks{e-mail: mabe@phys.titech.ac.jp}, \
        Akika Nakamichi\thanks{e-mail: akika@phys.titech.ac.jp} \
        and \ Tatsuya Ueno\thanks{e-mail: tatsuya@phys.titech.ac.jp}
        \\ \\ \\
        Department of Physics, Tokyo Institute of Technology \\
        Oh-okayama, Meguro-ku, Tokyo 152, Japan}

\maketitle

\vskip 3cm

\abstract{
 We propose a topological version of four-dimensional
(Euclidean) Einstein gravity, in which anti-self-dual 2-forms and
an SU(2) connection are used as fundamental fields.
 The theory describes the moduli space of conformally self-dual Einstein
manifolds.
 In the presence of a cosmological constant, we evaluate the index of
the elliptic complex associated with the moduli space.}

\thispagestyle{empty}

\newpage



 Topological gravity is a field theoretic description of the moduli
space of gravitational instantons.
 Such a theory was first considered by Witten \ \cite{witten1}, where
a topological version of conformal gravity in four dimensions was
presented as a gravitational analogue of topological Yang-Mills
theory (TYMT) \ \cite{witten2}.
 The moduli space of the theory, the space of conformally self-dual
gravitational instantons, was investigated in detail by Perry and
Teo \ \cite{perry}.
 Since the work of Witten there have been several attempts to construct
such four-dimensional topological theories modeling different
gravitational moduli spaces \ \cite{torre}-\cite{anselmi}.
 In Ref.\, \cite{lee}, we proposed a topological version of
four-dimensional (Euclidean) Einstein gravity with or without a
cosmological constant.
 This topological version is obtained by modifying an alternative
formulation of Einstein gravity developed by Capovilla et al. \
\cite{capovilla}, in which anti-self-dual 2-forms and an SU(2)
connection are used as fundamental fields, instead of the metric or the
tetrad.
 With an appropriate choice of gauge condition, the BRST invariant
quantum action of the topological theory becomes the classical Einstein
action plus ghost terms which cancel out all local degrees of freedom.
 However there still remain zero-modes in the quantum action.
 The (finite) number of them is closely related to the dimension of the
moduli space, which now consists of Einstein manifolds with self-dual
Weyl tensor.
 When the cosmological constant is non-zero, the moduli space is
up to orientation, identical with the one considered by Torre in
which the Weyl tensor is anti-self-dual \ \cite{torre}.
 In his paper the dimension of the moduli space is found to be zero
when the cosmological constant is positive, and the result is true
also in our case.
 In the case of non-zero cosmological constant, we evaluate the index
of an elliptic complex associated with our moduli space by applying the
Atiyah-Singer index theorem.
 We also discuss the case of vanishing cosmological constant and mention
the dimension of the moduli space on the $K3$ surface.
\par

\vskip 0.4cm

 We start with fundamental fields, a trio of 2-forms $\Sigma^k$,
which transform under the chiral local-Lorentz representation $(2,0)$ of
SU$(2)_L$$\times$SU$(2)_R$, and a connection 1-form $\omega^k$
associated with the SU$(2)_L$.
 It is shown that $\Sigma^k$ and $\omega^k$ are anti-self-dual with
respect to the SO(4) indices, when expressed with them instead of the
SU(2) indices $i,j,k, \cdots $, \ \cite{penrose}.
 We take $\Sigma^k$ as primary metric fields to define the metric
$g_{\mu \nu}$ on a four-dimensional oriented manifold $M_4$,
\footnote{Greek indices $\mu, \nu, \cdots,$ denote world indices on
$M_4$ while Latin indices $a,b, \cdots,$ the SO(4) indices.
 We use the notation for the SU(2) indices,
       $F \cdot G \equiv F^i G^i$ and
        $(F \times G)^i \equiv \varepsilon_{ijk} F^j G^k$, where
          $\varepsilon_{ijk}$ is the structure constant of SU(2).}
\begin{equation}
g^{1 \over 2} g_{\mu \nu}
           = -\frac{1}{12} \, {\epsilon^{\alpha \beta \gamma \delta}}
                             \, {\Sigma_{\mu \alpha}} \cdot
                             ({\Sigma_{\beta \gamma}} \times
                               {\Sigma_{\delta \nu}}) \ ,
            \qquad g \equiv det(g_{\mu \nu})\ .
                                                     \label{eq: three}
\end{equation}
 We propose the following action for topological gravity with
cosmological constant $\Lambda$ \ \cite{lee},
\begin{equation}
 S_{TG} = \int_{M_4} \Sigma^k \wedge F_k
  - \frac{\Lambda}{24} \Sigma^k \wedge \Sigma_k \, ,  \label{eq: one}
\end{equation}
 where $F^k \equiv d\omega^k + (\omega \times \omega)^k $.
 Varying the action with respect to each of fields $\Sigma^{k}$,
$\omega^{k}$, we obtain the equations of motion:
{
\setcounter{enumi}{\value{equation}}
\addtocounter{enumi}{1}
\setcounter{equation}{0}
\renewcommand{\theequation}{\theenumi\alph{equation}}
\begin{eqnarray}
  &&F^k - {\Lambda \over 12} \, \Sigma^k = 0  \ , \label{eqn: ins1}
\\
 D{\Sigma^k} =&& \, d{\Sigma^k} + 2\, (\omega \times \Sigma)^k = 0\ .
                                        \label{eqn: ins2}
\end{eqnarray}
\setcounter{equation}{\value{enumi}}
}
 For the $\Lambda  \not= 0$ case, eliminating $\Sigma^k$ from the action
by using (\ref{eqn: ins1}) we obtain the effective action proportional
to $\int F^k \wedge F_k $, which is the classical action of the TYMT for
the SU(2) gauge group \ \cite{baulieu}.
 On the other hand, for the $\Lambda = 0$ case, the action describes a
BF-type topological field theory \ \cite{blau},\cite{horowitz}.
 In this topological model, there exists a symmetry generated by a
parameter 1-form $\theta^k$ in addition to the local-Lorentz (with a
0-form $\phi^k$) and diffeomorphism (with a vector field $\xi^{\mu}$)
symmetries,
\begin{equation}
  \delta \omega^k
   =D\phi^k +{\tilde {\cal L}}_\xi \omega^k
    +{\Lambda \over 12}\theta^k  \ ,
\qquad
  \delta \Sigma^k
  =2(\Sigma \times \phi )^k +{\tilde {\cal L}}_\xi \Sigma^{k}
   +D\theta^k \ ,
                                                       \label{eq: four}
\end{equation}
 where we have used a modified action ${\tilde {\cal L}}_\xi$ for
diffeomorphism, which differs from usual Lie derivative ${\cal L}_\xi$
by the local-Lorentz transformation $\delta_{\tilde \phi}$ with
a parameter ${\tilde \phi}^k = \xi^{\mu} \omega^k_{\mu}$:
${\tilde {\cal L}}_\xi = {\cal L}_\xi - \delta_{\tilde \phi}$
\ \cite{myers}.
 The $\theta^k$-symmetry is regarded as a `restricted' topological
symmetry which preserves the equations of motion (3).
 With the appearance of the $\theta^k$-symmetry, the theory turns out
to be on-shell reducible in the sense that the transformation laws
(\ref{eq: four}) are invariant under
\begin{equation}
  \delta \phi^k = - \frac{\Lambda}{12} \epsilon^k \ ,
\qquad
  \delta \theta_{\mu}^k = D_{\mu} \epsilon^k
                      + 2  \rho^{\nu} \Sigma_{\nu \mu}^k \ ,
\qquad
    \delta \xi^{\mu} = - \rho^{\mu} \ ,
                                          \label{eq: five}
\end{equation}
as long as the equations of motion are satisfied.
 The transformations with parameters $\epsilon^k$ and $\rho^{\mu}$
correspond to redundant local-Lorentz and diffeomorphism symmetries
respectively.
\par

 Our strategy to construct a topological quantum field theory is to
consider the following equation as a gauge fixing condition for the
$\theta^k$-symmetry (except for the redundant part of the symmetry):
\begin{equation}
    {}^{tf.} \Sigma^i \wedge \Sigma^j
    \equiv \Sigma^{(i} \wedge \Sigma^{j)}
    - {1 \over 3} \delta_{ij} \Sigma^k \wedge \Sigma_k = 0 \ .
                                                 \label{eq: six}
\end{equation}
 This quadratic equation for $\Sigma^k$ is the constraint imposed in
original 2-form Einstein gravity \ \cite{capovilla} and is a necessary
and sufficient condition that $\Sigma^k$ is composed of tetrad 1-form:
$\Sigma^k (e) = - \eta^k_{ab} e^a \wedge e^b$, where $\eta^k_{ab}$ are
anti-self-dual constant coefficients called the t'Hooft's $\eta$
symbols \ \cite{t'hooft}.
 Such a 2-form $\Sigma^k (e)$ is anti-self-dual with respect to world
indices and via (\ref{eqn: ins1}), the curvature $F^k$ also becomes
anti-self-dual.
 The set of equations in (3),(\ref{eq: six}) arose before as
an ansatz within the framework of 2-form Einstein gravity with the
cosmological constant \ \cite{capovilla},\cite{samuel}.
 We consider them to be {\it instanton equations}.
 Here we assume a tensor-SU(2) connection $\nabla_{\mu}$ compatible with
$\Sigma^k$, that is, $\nabla_{\mu}\, \Sigma^k_{\nu \rho} = 0$, which
leads to the metricity condition.
 Then (\ref{eqn: ins2}) and (\ref{eq: six}) make the anti-self-dual part
of torsion vanish.
 We take a torsion-free extension to the self-dual part in order to
work on a Riemannian manifold.
 In the case we obtain the following relations between the $su(2)$
valued curvature $F^k$ and the anti-self-dual parts of Riemann curvature
tensor and Weyl tensor as follows:
\begin{equation}
{}^{(-)} R_{\mu \nu \rho \tau}= 4 F_{\mu \nu} \cdot
                         \Sigma_{\rho \tau} \ ,
\qquad \qquad \qquad
{}^{(-)} W_{\mu \nu \rho \tau}=
  4 \, ( F_{\mu \nu} -{\Lambda \over 12} \Sigma_{\mu \nu} )
     \cdot \Sigma_{\rho \tau} \ .
                                                 \label{eq: seven}
\end {equation}
 {}From (\ref{eqn: ins1}) and (\ref{eq: seven}), the Einstein equation
$R_{\mu \nu} = \Lambda g_{\mu \nu}$ and
${}^{(-)} W_{\mu \nu \rho \tau} = 0$ are derived.
 So our instanton equations determine conformally self-dual Einstein
manifolds.
 Note that in the case $\Lambda = 0$, the Ricci tensor becomes zero as
well as ${}^{(-)} W_{\mu \nu \rho \tau}$, or equivalently the Riemann
tensor is self-dual, ${}^{(-)} R_{\mu \nu \rho \tau}= 0$.
\par

 Since our topological model has on-shell reducible symmetries
(\ref{eq: five}), it is slightly complicated to quantize it by the
BRST gauge-fixing procedure, and so we adopt the
Batalin-Fradkin-Vilkovisky (BFV) formalism
\ \cite{batalin},\cite{henneaux}.
 In this canonical formalism we first obtain the nilpotent BRST charge
in the extended phase space.
 Using the BRST charge and gauge conditions for various symmetries in
the model, the partition function is defined as a path integral
with the BRST invariant quantum action
$S_q \equiv \int dt \, d^3 x [{\dot \varphi} \cdot {\cal P}_{\varphi}
- H_{eff}(\varphi, {\cal P}_{\varphi})]$, where $\varphi$ and
${\cal P}_{\varphi}$ are fields and their canonical momenta,
respectively and $H_{eff}$ is the BRST invariant effective Hamiltonian.
 Finally, integrating out conjugate momenta, we obtain the completely
covariant form of the quantum action $S_q$.
 It consists of the classical action $S_{TG}$ plus gauge fixing and ghost
terms, $\int \psi_{ij} \Sigma^i \wedge \Sigma^j + \cdots $, where the
term with a symmetric trace-free Lagrange multiplier field $\psi_{ij}$
comes from the gauge fixing term in (\ref{eq: six}).
 The sum $S_{TG} + \int \psi_{ij} \Sigma^i \wedge \Sigma^j$ is just the
original action of 2-form Einstein gravity \ \cite{capovilla}, and hence
the action $S_q$ becomes the classical Einstein action plus other terms
which cancel out local degrees of freedom.
 In $S_q$ there arise cubic (and higher than cubic) ghost terms, which
never emerge from the usual Faddeev-Popov procedure.
 This BRST quantization is reported in detail in \ \cite{abe}.
\par

\vskip 0.4cm

 To investigate the number of zero-modes in the quantum action $S_q$, we
consider the moduli space ${\cal M}$ defined by our instanton equations
(3),(\ref{eq: six}) describing conformally self-dual Einstein manifolds.
 Given a solution $(\Sigma^k_0, \omega^k_0)$ of the instanton equations,
the tangent space $T({\cal M})$ of ${\cal M}$ at the point is identical
with the space of infinitesimal fluctuations
$(\delta \Sigma^k,\, \delta \omega^k)$ which satisfy linearized
instanton equations,
\begin{equation}
 D_1 (\dS, \dO) \equiv
     (\,D \dO -{\Lambda \over 12} \delta \Sigma^k \, ; \
     D \dS + 2 (\delta \omega \times \Sigma_0)^k \, ; \
     2 \, {}^{tf.} \Sigma^i_0 \wedge \delta \Sigma^j \, ) = 0 \ ,
                                           \label{eq: nine}
\end{equation}
modulo fluctuations generated by the local-Lorentz (SU(2))
transformation and diffeomorphism:
\begin{equation}
  T({\cal M})_{(\Sigma^k_0, \omega^k_0)}
  =  \{(\dS, \dO) \vert D_1 (\dS, \dO)= 0 \} /
            \{ SU(2) \times diffeo. \}  \ .
                                                  \label{eq: ten}
\end{equation}
 We define the following sequence of mappings on a compact conformally
self-dual Einstein manifold:
\begin{eqnarray}
  && \qquad \Omega_0 \qquad \qquad \qquad \ \ \Omega_1
     \qquad \qquad \qquad \qquad  \ \ \Omega_2  \nonumber
\\
 0 \mathrel{\mathop \rightarrow^{D_{-1}}} \,
      &&{\left \{ \matrix{\hbox{\rm space of}\hfill \cr
        \hbox {$(\phi^k, \xi^{\mu})$}\hfill \cr} \right \}} \,
   \mathrel{\mathop \rightarrow^{D_0}} \,
        {\left \{ \matrix{\hbox{\rm space of}\hfill \cr
         \hbox{$(\dS,\, \dO)$} \hfill \cr} \right \}}      \,
   \mathrel{\mathop \rightarrow^{D_1}} \,
        {\left \{ \matrix{\hbox{\rm space of}\hfill \cr \hbox
       {\rm instanton equations}\hfill \cr} \right \}}   \,
   \mathrel{\mathop \rightarrow^{D_2}} \, 0 \, .
                                           \label{eq: eleven}
\end{eqnarray}
 In the above sequence $D_{-1}$ and $D_{2}$ are identically zero
operators.
 The operator $D_0$ is defined by
\begin{equation}
  D_0 (\phi^k, \xi^{\mu}) \equiv
      (\, 2(\Sigma_0 \times \phi )^k
       +{\tilde {\cal L}}_\xi \Sigma^k_0 \, ; \
       D\phi^k +{\tilde {\cal L}}_\xi \omega^k_0 \,)   \ .
                                             \label{eq: twelve}
\end{equation}
 We can easily check that $D_1 \, D_0 = 0$ and the ellipticity of the
sequence (\ref{eq: eleven}).
 Hence it is an {\it elliptic complex}.
 Defining the inner product in each space $\Omega_i$, we can introduce
the adjoint operators $D_0^*$ and $D_1^*$ for $D_0^{}$
and $D_1^{}$ respectively and the Laplacians $\triangle_i$;
$ \triangle_0 = D_0^* D_0^{} \, , \
  \triangle_1 = D_0^{} D_0^* + D_1^* D_1^{} \, , \
  \triangle_2 = D_1^{} D_1^* \,$.
 We may then define the cohomology group on each $\Omega_i$,
\begin{equation}
 H^i \equiv {\rm Ker}\, D_i/{\rm Im}\, D_{i-1} \ .
                                                \label{eq: thirteen}
\end{equation}
 It is easy to show that $H^i$ is equivalent to the kernel of
$\triangle_i$, the harmonic subspace of $\Omega_i$.
 These cohomology groups are finite-dimensional and we set
$h^i \equiv dim. H^i$.
 The $H^1$ is exactly identical with the tangent space of ${\cal M}$ in
(\ref{eq: ten}), the dimension of which we need to know.
 On the space $\Omega_0$, $H^0$ is equal to ${\rm Ker}\, D_0$ because
the image of $D_{-1}$ is trivial.
 In the $\Lambda \not = 0$ case, Torre found that ${\rm Ker}\, D_0$ is
equivalent to the space of the Killing vectors \ \cite{torre} and
this is also true in the $\Lambda = 0$ case.
 The kernel of $D_2$ is the whole of the space $\Omega_2$.
 Hence $H^2$ is the subspace of $\Omega_2$ orthogonal to the mapping
$D_1$, or equivalently it is the kernel of $D_1^*$.
 The index of the elliptic complex is defined as the alternating sum,
\begin{equation}
  {\rm Index} \equiv h^0 - h^1 + h^2 \ .
                                              \label{eq: fourteen}
\end{equation}
 This is a topological quantity determined by the Atiyah-Singer index
theorem \ \cite{shanahan}.
 In the $\Lambda \not= 0$ case, the elliptic complex can be reduced to a
more simple form by restricting $\Omega_1$, the space of $(\dS, \dO)$,
to the subspace ${\tilde \Omega_1}$ in which the first linearized
equation $D \dO -{\Lambda \over 12} \delta \Sigma^k = 0$ in
(\ref{eq: nine}) is satisfied.
 Then $\delta \Sigma^k$ are linearly dependent on $\delta \omega^k$
and ${\tilde \Omega_1}$ becomes the space of $\delta \omega^k$.
 Besides the second linearized equation becomes trivial on
${\tilde \Omega_1}$ since it is derived from the first equation by
operating the covariant derivative $D$ to it.
 Eliminating $\Sigma^k$ from the third linearized equation in
(\ref{eq: nine}) we define the operator ${\tilde D_1}$,
\begin{equation}
 {\tilde D_1}(\dO) \equiv D_1 (\dS, \dO) \vert_{{\tilde \Omega_1}} \,
= 2 \, {}^{tf.} \Sigma^i_0 \wedge \delta \Sigma^j \,
= 2\, (12/\Lambda)^2 \, {}^{tf.} F^i_0 \wedge D \, \delta \omega^j \ .
                                                    \label{eq: newone}
\end{equation}
 To elucidate each of spaces in (\ref{eq: eleven}), we introduce
general spin bundles $\Omega^{m,n}$, the space of fields with spin
$(m,n)$ of SU(2) $\times$ SU(2) \ \cite{penrose}:
the space ${\tilde \Omega_1}$ of $su(2)$ valued 1-forms $\dO$ is
$\Omega^{2,0} \otimes \Lambda^1 \simeq \Omega^{2,0}
\otimes \Omega^{1,1}$ while the space $\Omega_0$ of $(\phi^k,\xi^{\mu})$
is equivalent to $\Omega^{2,0} \oplus \Lambda^1 \simeq \Omega^{2,0}
\oplus \Omega^{1,1}$.
 The elements of the image of ${\tilde D_1}$ are 4-forms with symmetric
trace-free SU(2) indices, that is, with the spin (4,0).
 Hence the space $\Omega_2$ becomes
$\Omega^{4,0} \otimes \Lambda^4 \simeq \Omega^{4,0}$.
 In terms of these spin bundles, the elliptic complex (\ref{eq: eleven})
is written as
\footnote{The space $\Omega_0$ in the complex of Torre is equivalent to
$\Omega^{2,0} \oplus \Omega^{2,0} \oplus \Omega^{0,0}$, which
seems slightly different from ours, $\Omega^{2,0} \oplus \Omega^{1,1}$
up to harmonic parts of  the former.
 This leads to different results of the index.}
\begin{eqnarray}
  && \qquad \Omega_0 \qquad \qquad \quad \ \ {\tilde \Omega_1}
     \qquad \qquad \Omega_2  \nonumber
\\
 0 \stackrel{D_{-1}} \to \ &&\Omega^{2, 0} \oplus \Omega^{1, 1} \
   \stackrel{D_0} \to \ \Omega^{2, 0} \otimes \Omega^{1, 1} \
   \stackrel{{\tilde D_1}}{\to} \ \Omega^{4, 0} \
   \stackrel{D_2} \to \ 0 \ .
                                             \label{eq: fifteen}
\end{eqnarray}
 By applying the Atiyah-Singer index theorem to the elliptic complex,
we obtain
\begin{equation}
 {\rm Index} = \int_{M_4}
 \frac{{\rm ch} ( \Omega^{2,0} \oplus  \Omega^{1,1} \, \ominus \,
                  \Omega^{2,0} \otimes \Omega^{1,1} \, \oplus \,
                  \Omega^{4,0} )\,
      {\rm td} (TM_4 \otimes {\bf C} ) }
      {{\rm e} (TM_4) } \\               = 5 \chi - 7 \tau \, ,
                                             \label{eq: sixteen}
\end{equation}
where ${\rm ch},$ ${\rm e}$ and ${\rm td}$ are the Chern
character, Euler class and Todd class of the various vector bundles
involved.
 Therefore the alternating sum of $h^i$ in (\ref{eq: fourteen}) is
determined by the Euler number $\chi$ and Hirzebruch signature $\tau$.
 By changing $\tau \rightarrow \mid \tau \mid $, this index can be
also adopted to manifolds with the opposite orientation.
 If $\Lambda > 0$, as shown by Torre \ \cite{torre}, $h^1$ and $h^2$ are
found to be zero.
 Therefore from (\ref{eq: fourteen}) and (\ref{eq: sixteen}), the
dimension $h^0$ is equal to the index,
\begin{equation}
  h^0 = 5 \chi - 7 \tau \ ,
\qquad
  h^1 = h^2 = 0 \ .
                                             \label{eq: seventeen}
\end{equation}
 The value of $h^0$, the dimension of the Killing vector space, agrees
that obtained by a different method in \ \cite{besse}.
 On the contrary if $\Lambda < 0$, $h^0$ becomes zero \ \cite{torre},
although $h^1$ and $h^2$ are not completely determined:
\begin{equation}
 h^0 = 0 \ ,
\qquad
 h^2-h^1 = 5 \chi - 7 \tau \ .
                                              \label{eq: eighteen}
\end{equation}
 Well known examples of compact conformally self-dual Einstein manifolds
with positive cosmological constant are the four-sphere $S^4$ with the
standard metric, and the complex projective two-space $CP^2$ with the
Fubini-Study metric.
 {}From (\ref{eq: seventeen}), $\, h^0$ for $S^4$ and $CP^2$ are 10 and 8
respectively.
 These examples are important because of the following theorem of
Hitchin \ \cite{besse}. \par
 Let $M_4$ be a compact conformally self-dual Einstein manifold. Then
\\
(i) If $R > 0 \ (\Lambda > 0)$, $M_4$ is either isometric to $S^4$, or
to $CP^2$, with their standard metrics; \\
(ii) If $R = 0 \ (\Lambda = 0)$, $M_4$ is either flat or its
universal covering is the $K3$ surface with the Calabi-Yau metric.
\par
 In the $\Lambda = 0$ case, the SU(2) spin connection $\omega^k$ can be
gauged away when $M_4$ is simply connected and then instanton equations
reduce to
\begin{equation}
      d\Sigma^k=0 \ ,
\qquad \qquad
      {}^{tf.} \Sigma^i \wedge \Sigma^j =0 \ .
                                               \label{eq: nineteen}
\end{equation}
 These equations give the Ricci-flat condition and determine $\Sigma^k$
to be a trio of closed K\"ahler forms
\ \cite{kunitomo},\cite{capovilla}.
 In Ref. \ \cite{plebanski}, Plebanski used these equations to
derive his `heavenly equations'.
 A manifold which satisfies (\ref{eq: nineteen}) is called
hyper-K\"ahlerian.
 The $K3$ surface is a unique simply connected compact manifold with
such a hyper-K\"ahler structure.
 The elliptic complex (\ref{eq: eleven}) now describes the
deformation of the three K\"ahler forms and it can be decomposed to
deformations of metric and complex structures.
 A careful examination of the intersection part of these deformations
reveals that the dimension of the moduli space of K\"ahler forms
on the $K3$ surface is 59 \ \cite{besse}.

\vskip 0.4cm


 In this paper, we have presented a topological version of 2-form
Einstein gravity in four dimensions.
 For a compact manifold, we have defined the elliptic complex associated
with the moduli space of our theory.
 By applying the Atiyah-Singer index theorem in the $\Lambda \not= 0$
case, we have evaluated the index of the elliptic complex. \par

 When the dimension of the moduli space is non-zero, as occurs when
$\Lambda = 0$, there arise fermionic zero-modes the number of which is
equal to the dimension, and these make the partition function trivial.
 To avoid this we need some functional ${\cal O}$ which absorbs the
zero-modes.
 If one calculates the vacuum expectation value of the `observable'
${\cal O}$, then it may provide non-trivial information such as a
differential invariant to distinguish differential structures on
conformally self-dual Einstein manifolds.
 Such a functional ${\cal O}$ is required to be BRST invariant to
preserve the topological nature of the theory and may be obtained from
the BRST descendant equations as in two-dimensional topological
gravity \ \cite{witten3}. \par

 It would be intriguing to study the $\Lambda =0$ case since the
relation of four-dimensional (Riemann) self-dual gravity and
two-dimensional conformal field theory has been investigated.
 In fact, Park showed that the former arises from a large N limit
of the two-dimensional sigma model with SU(N) Wess-Zumino terms
only \ \cite{park}.
 Our topological model will be useful to understand the relation and
to develop the self-dual gravity.

\vskip 2.0cm

 \ We are grateful to Q-Han Park and N. Sakai for useful discussions.
 We also acknowledge P. Crehan for reading the manuscript.



\end{document}